# Niels Bohr on the wave function and the classical/quantum divide[1]

Henrik Zinkernagel
Department of Philosophy I
University of Granada, Spain.
zink@ugr.es

**Abstract**

It is well known that Niels Bohr insisted on the necessity of classical concepts in the account of quantum phenomena. But there is little consensus concerning his reasons, and what he exactly meant by this. In this paper, I re-examine Bohr's interpretation of quantum mechanics, and argue that the necessity of the classical can be seen as part of his response to the measurement problem. More generally, I attempt to clarify Bohr's view on the classical/quantum divide, arguing that the relation between the two theories is that of mutual dependence. An important element in this clarification consists in distinguishing Bohr's idea of the wave function as symbolic from both a purely epistemic and an ontological interpretation. Together with new evidence concerning Bohr's conception of the wave function collapse, this sets his interpretation apart from both standard versions of the Copenhagen interpretation, and from some of the reconstructions of his view found in the literature. I conclude with a few remarks on how Bohr's ideas make much sense also when modern developments in quantum gravity and early universe cosmology are taken into account.

## 1. Introduction

Foundational discussions of quantum mechanics routinely include reference to the Copenhagen interpretation. Though this interpretation is supposed to originate with Niels Bohr, there is often some confusion concerning both what the Copenhagen interpretation precisely amounts to, and which parts of this interpretation are associated with whom of the founding figures of quantum mechanics. But there are signs that the mist is beginning to clear up. For instance, Howard (2004) convincingly argues that critics of the Copenhagen interpretation often conflate Heisenberg's views with Bohr's, and Faye (2008) gives a very helpful overview of the main tenets in Bohr's thinking.

Even among friends of Bohr, however, there are still disagreements about how best to understand him. It is well known that Niels Bohr insisted on the necessity of the concepts of classical physics in the description of quantum phenomena. But there is little consensus concerning the justification, and the philosophical implications of this idea. Relatedly, or so I will argue, there is little consensus concerning Bohr's view on the wave function, the quantum measurement problem, and, more generally, the relation between classical and quantum physics. This paper is part of a series which aim to clarify and defend Bohr's interpretation of quantum mechanics. At the same time, the proposed clarification should be helpful for situating Bohr in relation to contemporary philosophical debates of quantum mechanics in which the wave function, the measurement problem and the classical/quantum relationship are central topics.

---

[1] Published in *Studies in History and Philosophy of Modern Physics*, Vol. 53, 2016, pp. 9-19.

The lack of consensus among Bohr commentators regarding the role and status of the classical concepts is perhaps not surprising, given Bohr's own enigmatic accounts of his view. In one of its most quoted forms, Bohr expressed the necessity of classical concepts thus (1949, p. 39):

> [I]t is decisive to recognize that, *however far the phenomena transcend the scope of classical physical explanation, the account of all evidence must be expressed in classical terms.* The argument is simply that by the word "experiment" we refer to a situation where we can tell others what we have done and what we have learned and that, therefore, the account of the experimental arrangement and of the results of the observations must be expressed in unambiguous language with suitable application of the terminology of classical physics. [Emphasis in original]

Generations of physicists and philosophers have questioned the reasoning behind, and the simplicity of, this type of argument. For instance, after having read Bohr's response to the famous Einstein, Podolsky and Rosen paper, Schrödinger wrote in a letter to Bohr:

> There must be quite definite and clear grounds, why you repeatedly declare that one *must* interpret observations classically, which lie absolutely in their essence…. It must belong to your deepest conviction and I cannot understand on what you base it.
> [Schrödinger to Bohr (1935), quoted from Howard (1994, p. 201)]

In what follows I will attempt to answer Schrödinger's question on behalf of Bohr by re-examining Bohr's writings.[2] This will include, in section 2, briefly recalling Bohr's notion of complementarity, clarifying his view on the wave function, and providing new evidence regarding his conception of the wave function collapse. With these elements in place, section 3 will frame the necessity of the classical in terms of Bohr's response to the measurement problem, and his demand for a reference frame. Section 4 will treat the more general question of the relation between the classical and the quantum in light of Bohr's view on the stability of matter (an issue which has so far been little discussed in the literature). Finally, in section 5, I sum up and briefly indicate how Bohr's interpretation of quantum mechanics still makes much sense also in the current landscape of cosmology and fundamental physics.

## 2. Elements of Bohr's view

### 2.1 Complementarity

Bohr's idea of complementarity has been treated extensively in the literature. But it is worthwhile to give a brief summary of the idea since, as we shall see below, it provides a good starting point regarding Bohr's view on the wave function and wave function collapse. In general, complementarity means that the attribution of certain properties to quantum objects can take place only in experimental contexts which are mutually incompatible. Thus, for example, an experiment which can determine the position of an electron cannot be used to determine its momentum. Complementary properties, such as position and momentum, are both necessary for a full understanding of the object but, as manifested in Heisenberg's uncertainty relations, the object cannot possibly be attributed precise values of both properties at the same time.[3]

---

[2] In the original 1935 response to Schrödinger (which can be found in Kalckar 1996, p. 511), Bohr points out that the description of the measurement set-up must "…involve the arrangement of the instruments in space and their functioning in time, if we shall be able to state anything at all about the phenomena". Bohr then argues that the measurement apparatus, in order to serve as such, must be kept outside the system to which quantum mechanics is applied. While this short answer was probably insufficient to satisfy Schrödinger's demand for an explanation, I will attempt to unpack and defend it below.

In a characteristic exposition from 1949, Bohr introduces the notion of complementarity by considering two variations of the famous double slit experiment. In the first, the diaphragms are kept fixed, and this allows for the appearance of an interference pattern on the photographic plate. In the second variation, the single-slit diaphragm is suspended by a spring, and so is allowed to move vertically (fig. 1).

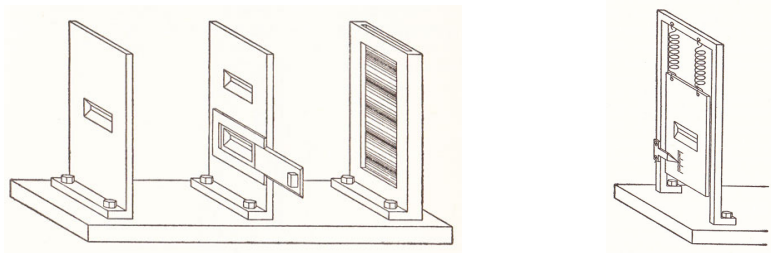

Figure 1: Two variations of the double slit experiment. In the second variation, only the spring-suspended single slit diaphragm is depicted. Taken from Bohr (1949, p. 48).

The movable diaphragm permits control of its momentum before and after the passage of the particle, and therefore the determination of which of the double slits the particle subsequently moves through. However, Bohr (1949, p. 46) observes that "…we are presented with a choice of *either* tracing the path of a particle *or* observing interference effects" [emphasis in original]. The point is that *if* the path is tracked, e.g. by controlling the momentum gain of the spring-suspended diaphragm due to the particle's passage, the position of this diaphragm when the particle passes through becomes uncertain. The uncertainty is the result of an uncontrollable interaction (involving e.g. a momentum change of the diaphragm) when the momentum measurement is made, and it implies a washout of the interference pattern.[4] Bohr concludes (1949, 46):

---

[3] The relation between complementarity and the uncertainty relations was spelled out already in Bohr's Como lecture (1928, p. 60): "According to the quantum theory a general reciprocal relation exists between the maximum sharpness of definition of the space-time and energy-momentum vectors associated with the individuals. This circumstance may be regarded as a simple symbolical expression for the complementary nature of the space-time description and the claims of causality". For the purposes of this paper, I bypass the controversial issue of whether, and to which extent, Bohr's views on complementarity changed over the years; see e.g. discussion and references in Faye and Folse (1998).

[4] This may suggest that the momentum measurement disturbs some pre-existing definite position of the diaphragm (for instance, Fine and Beller (1994, p. 13) read Bohr this way). However, in this experimental context, Bohr took the particle and the diaphragm to be described by a non-separable (or entangled, see below) state *until* the momentum control is carried out (see also Bohr 1938, p. 102). For in this case, the diaphragm is part of the system to which quantum mechanics should be applied: "As regards the

> We have here to do with a typical example of how the *complementary phenomena* appear under mutually exclusive experimental arrangements... and [we] are just faced with the *impossibility*, in the analysis of quantum effects, of drawing any sharp separation between an independent behaviour of atomic objects and their interaction with the measuring instruments which serve to define the conditions under which the phenomena occur. [My emphasis]

As we shall see below, one way this impossibility is manifested is through Bohr's idea that the wave function (e.g. of an electron in a double slit experiment) only refers to the object in a given experimental context. Thus, for instance, insofar as the electron can be described as a traveling wave in the double slit experiment, this description only makes sense in a setup in which we plan to investigate the interference pattern.

Furthermore, Bohr associates complementarity with what he calls indivisibility which, as we shall see, is closely related to his notion of wave function collapse. Bohr writes (1956, p. 168):

> The essential indivisibility of proper quantum phenomena finds logical expression in the circumstance that any attempt at a well-defined subdivision would require a change in the experimental arrangement that precludes the appearance of the phenomenon itself. Under these conditions, it is not surprising that phenomena observed with different experimental arrangements appear to be contradictory when it is attempted to combine them in a single picture. Such phenomena may appropriately be termed complementary in the sense that they represent equally important aspects of the knowledge obtainable regarding the atomic objects and only together exhaust this knowledge.

The indivisibility of quantum phenomena thus implies that the precise measurement process (i.e. what is physically going on when a quantum object interacts with an instrument) is unanalyzable in terms of a causal space-time description. This was a main point already in the Como lecture (Bohr 1928), in which the indivisibility is an expression of what Bohr called the quantum postulate.

Indivisibility also means that object and apparatus are dynamically inseparable until a measurement has been made; see also Faye (2008). Here is what Bohr says in the Como lecture (1928, p. 54):

> Now the quantum postulate implies that any observation of atomic phenomena will involve an interaction with the agency of observation not to be neglected. Accordingly, an *independent* reality in the ordinary physical sense can neither be ascribed to the phenomena nor to the agencies of observation. [My emphasis]

In his interesting and influential reconstruction of Bohr, Howard (1994, 2004) interprets this point in terms of entanglement between the object and the apparatus. I believe this is consistent with what Bohr writes elsewhere, as long as it is taken to mean that *parts*

---

quantum-mechanical description, we have to deal here with a two-body system consisting of the diaphragm as well as of the particle, ..." (Bohr 1949, p. 45). Hence it is the *combined* system of object and diaphragm (and not the diaphragm alone) which is 'disturbed' by the momentum measurement.

of the apparatus (e.g. the spring-suspended diaphragm from above) are entangled with the object. But a main point of the present paper is that Bohr could not have subscribed to the idea that the apparatus, *as a whole*, is entangled with the object. As will be discussed further below, the reason is that the apparatus must also encompass classically described parts.

## 2.2 The wave function

A key to the difference between Bohr's interpretation and various versions of what has become known as the Copenhagen interpretation lies in the understanding of the wave function. Moreover, as I shall argue below, Bohr's understanding of the wave function was not quite what even careful commentators have taken it to be.

For a start, discussions of the Copenhagen interpretation in the literature are ambiguous between two different views of the wave function, both of which of course accept the Born interpretation.[5] Sometimes the Copenhagen (and Bohr's) interpretation is associated with the *epistemic* view of the quantum state, according to which the quantum state is but a representation of our knowledge of the physical system, and thus not a real existing entity in itself. On this view the 'collapse' of the wave function is not a physical process, and it just reflects an update of our information about the system; see e.g. Zeilinger (1999). By contrast, the Copenhagen interpretation has also been associated with an *ontological* view of the quantum state, in which the wave function somehow describes a real wave, and the collapse is a real physical process – presumably induced by the observer. This ontological view is usually attributed to von Neumann in his 1932 textbook exposition of quantum mechanics; see e.g. Henderson (2010).

Bohr scholars have mostly associated his interpretation of quantum mechanics with a rejection of the ontological view of the quantum state. For instance, Murdoch (1987, p. 122) argues that Bohr preferred the epistemic interpretation; also as opposed to the statistical interpretation in which the wave function describes only ensembles of individual microphysical objects. Likewise, Faye (2008) asserts that Bohr held an instrumental view in which "…the state vector's representational function should not be taken literally but be considered a tool for the calculation of probabilities of observables". However, while it is true that Bohr mostly stressed epistemological and instrumental aspects of quantum mechanics (e.g. by speaking of the *description* of quantum systems, and of the theory as a *tool* for predictions), he also hinted that the wave function was not merely epistemic (or instrumental or statistical). The fact is that the wave function, for Bohr, is *symbolic* – a symbolic representation of a quantum system – and his view appears to have certain ontological implications regarding representation and explanation.[6] To see this, let us first look at why Bohr did not accept a straightforward ontological reading of the wave function as a real wave.

In the Como lecture Bohr gives three reasons for why the wave function cannot be visualized (and interpreted) as a real wave in space-time: i) The wave function is imaginary, that is, its mathematical expression contains the square root of -1; ii) it is formulated in configuration space which, in general, has more than the usual three spatial dimensions; and iii) "Schrödinger's formulation of the interaction problem… involves a neglect of the finite velocity of propagation of forces claimed by relativity theory" (i.e. the Schrödinger equation is non-relativistic), Bohr (1928, p. 77). Bohr stresses the second reason as the more important one for this non-visualizability, which excludes an ontological reading of the wave function as representing a classical wave in space-time. As Bohr puts it in 1932, "…the symbolic aspect of Schrodinger's wave functions appears immediately from the use of a multidimensional co-ordinate space, essential for their representation in the case of atomic systems with several electrons" (Kalckar 1985, p. 401).

---

[5] According to the Born interpretation, the absolute square of the wave function gives the probabilities of measuring the different possible values of the observables of a quantum system (e.g. the position of a particle).
[6] For an interesting analysis of Bohr's notion of 'symbolic' in light of post-Kantian philosophy, see Chevalley (1996, pp. 240-242) and references therein.

Thus, although the quantum state represents the system or object, it cannot – in general – be given a pictorial interpretation (i.e. it is not visualizable as a real wave in spacetime). However, this also means that there can be *special* situations, like the free non-relativistic electron in the double slit experiment with fixed diaphragms (where the wave function is three-dimensional), in which we *can* visualize or picture the development of (at least the square of) the wave function as a representation of a quantum object moving through the experimental arrangement. In such cases the wave function can be identified with the wave field (or wave packet) associated with the motion of a single particle.[7] This identification can be seen, for instance, in the response to Einstein, Podolsky and Rosen, where Bohr notes regarding the case of a particle passing through a slit in a diaphragm (1935, p. 75): "…the diffraction by the slit of the plane wave giving the symbolic representation of its [the particle's] state [as a spherical wave] will imply an uncertainty in the momentum of the particle, after it has passed the diaphragm…" [my inserts]. A similar statement can be found in Bohr (1949, p. 43), where Bohr also includes the traveling wave fronts in a pictorial representation of the double slit experiment (1949, p. 45, figure 3).

Bohr's account of the interference pattern in the double slit experiment, in terms of traveling wave fronts and diffraction of plane waves, strongly suggests that the wave function in this case has an explanatory role which goes beyond an instrumentalist reading of this function (insofar as it represents the electron as a wave field going through both slits). The reliance on wave aspects of the electron in the account of the interference pattern is also apparent when Bohr says (1949, p. 45): "With intense beams, this pattern is built up by the accumulation of a large number of individual processes, each giving rise to a small spot on the photographic plate, and the distribution of these spots follows a simple law *derivable from the wave analysis*" (my emphasis). This explanation (using the wave aspect of the particle's motion) only accounts for the *distribution* of the spots and not for the fact that the spots appear in the first place (in accordance with the idea that the precise measurement process is unanalyzable). But it implies that Bohr's conception of the wave function is not vulnerable to the critique, indicated e.g. by Pusey et al. (2012), that a purely epistemic or instrumental interpretation of the wave function appears insufficient to explain e.g. the wave-like interference pattern in the double slit experiment. In any case, even for a single electron wave function, a full ontological interpretation as a real wave in spacetime is excluded by it being complex (and non-relativistic).

A further important element in Bohr's conception of the symbolic (as opposed to ontologically real) wave function is that it is contextual. Here is what Bohr says (1958, p. 5), after having explained the necessity of a classical physical description of the measuring device:

---

[7] For an animation of a wave packet (its absolute square and its real/imaginary parts) moving through the double slit experiment, see e.g. Postnikov and Loktionov (2013).

> In the treatment of atomic problems, actual calculations are most conveniently carried out with the help of a Schrödinger state function, from which the statistical laws governing observations obtainable under specified conditions can be deduced by definite mathematical operations. It must be recognized, however, that we are here dealing with a purely symbolic procedure, the unambiguous physical interpretation of which in the last resort requires a reference to a complete experimental arrangement.

Thus, for Bohr, *the wave function is a representation of a quantum system in a particular, classically described, experimental context.* Three important points need to be made regarding this contextuality: 1) When a measurement is performed (that is, when an irreversible recording has been made; see below), then the context changes, and hence the wave function changes. This can formally be seen as a "collapse" of the wave function, with the square quotes indicating that we are *not* talking about a physical process in which a real wave collapses. 2) The distinction between an epistemic and ontic view of the wave function is sometimes taken to be that between a "representation of an agent's knowledge of the system" and a "representation of the system" (see e.g. Friederich 2013). In this sense, Bohr's view of the wave function is ontic, for – given an experimental context – the wave function is not agent-relative (i.e. the wave function does not depend on what a particular agent knows about the system but on the experimental arrangement and what in fact has happened to the system, e.g. whether or not the electron has been registered). 3) The experimental context is classically described (see section 3), which implies that this context is omitted from the quantum description, and thus not represented by means of a wave function. This furthermore implies that no wave function can be ascribed to the measurement apparatus as a whole (including recording devices such as pointers or photographic plates). For this reason, the above mentioned reconstruction of Bohr in terms of entanglement between object and apparatus (Howard 1994, 2004), cannot work if applied to the measuring apparatus as a whole; see also below and Zinkernagel (2015).[8]

## 2.3 Wave function collapse

Among Bohr scholars it is common to assert that Bohr never mentions the wave function collapse (see e.g. Howard 2004 and Faye 2008). It is true that in Bohr's published writings, he does not discuss the status or existence of this standard component in the popular image of the Copenhagen interpretation. However, in his unpublished works and letters one can find at least a few occasions where the collapse is mentioned. In a draft version of the Como lecture from October 12, 1927, Bohr accepts the idea of wave packet collapse (introduced in Heisenberg's 1927 paper on the uncertainty relations), and he links it to the statistical character of quantum mechanics (Kalckar 1985, p. 94):

> Due to the gradual spreading of the wave fields associated with the individuals the statistical character of the description, however, is by no means limited to the inaccuracy expressed by Heisenberg's relations. Indeed, we are forced to contemplate a proper reduction of the spatial extension of the fields after every new observation.

---

[8] Gomatam (2007) rejects that Bohr subscribed to a contextual reading of the wave function, and argues instead that Bohr's wave function describes both the object and the (whole) measurement apparatus. However, Gomatam's interpretation of Bohr is based on a misreading of a central Bohr quote. The relevant quote is from Bohr (1949), and is reproduced in Gomatam's text (2007, 6) like this: "The main point here is the distinction between the objects under investigation and the measurement instruments which serve to define, in classical terms, the conditions under which the phenomena appear . . . *these bodies together with the particles would in such a case constitute the system to which the quantum mechanical formalism is to be applied.*" [Gomatam's emphasis]. Nevertheless, in the omitted text (indicated by "…") Bohr makes clear that "these bodies" refer to a case in which *parts* of the measurement apparatus (like the suspended diaphragm in the double slit experiment) are included in the quantum system, and that this inclusion is "…in contrast to the proper measuring instruments" (Bohr 1949, p. 50). Consequently, these 'proper measuring instruments' are not described by a wave function. This moreover means that they cannot be described as being 'entangled' with the object.

At the outset of the Como lecture (both drafts and final version), Bohr had associated the statistical character of the quantum formalism with the lack of a rigorous definition of the object system, due to the unavoidable interaction with the measuring instruments (this interaction being implied by the quantum postulate and formally expressed through Heisenberg's relations). But in the draft quoted here, Bohr now points out that an additional statistical element, due to a "proper reduction" or collapse of the wave fields, is associated with each observation or measurement. In spite of this, the reduction or collapse does not appear in the published version of the Como lecture. What changed Bohr's mind? One can only speculate, but it seems reasonable to assert that it has to do with the intervention of Bohr's close friend Wolfgang Pauli. Indeed, Bohr urged Pauli for comments to a draft version of the Como lecture.[9]

In the beginning of his response to Bohr's query, Pauli first points out with respect to the origin of the statistical character of the quantum description (Kalckar 1985, p. 32):

> …it has become clear to me that the statistical interpretation of the theoretical results always enters at the point when one divides a closed system into two parts, which one interprets as object under observation and measuring instrument, respectively, and then asks what one can say about one part without knowledge of the other.
>
> (Pauli to Bohr, October 17, 1927)

A bit further down in Pauli's response, he comments directly on Bohr's mentioning of the "proper [or discontinuous][10] reduction of the spatial extension of the fields" (Kalckar 1985, p. 33):

> This is of course just a point which was not quite satisfactory in the Heisenberg paper; there the 'reduction of the wavepacket' seemed a bit mysterious. Now it should of course be emphasized that such reductions first of all are not necessary when all the measuring instruments are included in the system. [But] in order to be able to describe the results of

---

[9] In a letter from October 11, 1927, Bohr tells Pauli that "…I would be very grateful if you would immediately send me the proofs with all critical remarks which comes to your mind" [my translation from the Danish], Hermann and Von Meyenn (1979, p. 410). The draft sent to Pauli is presumably nearly identical to the one quoted above; see Kalckar (1985, p. 30).

[10] As Kalckar 1985, p. 33, footnote 34 explains, a correct translation of "unstetigen" appearing in the original German text of Pauli's letter (quoting a lost German version of Bohr's manuscript) should be "discontinuous" instead of "proper".

> observation theoretically at all, one must ask what can be said about one *part* of the total system on its own. And then one sees as a matter of course the complete solution – that the omission of the instruments of observation in many cases (not always, of course) may formally be replaced by such discontinuous reductions."[11]
>
> (Pauli to Bohr, October 17, 1927, emphasis in original)

Pauli's suggestion appears to correspond to the idea, in modern language, of taking a "partial trace" of a composite entangled state to obtain a reduced state for one of several interacting subsystems. In the case where one is interested only in finding out about the object, this reduced state will be a statistical mixture of collapsed object states. However, as often pointed out, the reduced state (or reduced density matrix) is not "ignorance interpretable" (the mixture is improper) and so it cannot be interpreted as if the object were actually in an unknown definite state. Taking a partial trace is therefore not equivalent to a collapse of the wave function; see e.g. Schlosshauer (2007, p. 333).[12] Hence, if Pauli is thinking along the lines of reduced states, his assertion that the collapse or reduction of the wave function is merely a formal replacement for the omission (or "tracing out") of the instrument is problematic. In any case, if this reduced state reading of Pauli is correct, he would seem to treat both object and instrument as (entangled) quantum systems.[13]

It would have been very interesting to have a written response from Bohr to Pauli regarding the collapse. But whereas Bohr may have received Pauli's letter just before leaving for the famous 1927 Solvay conference (which took place from 24-29 October), it is natural to think that the two men discussed the matter in person at that meeting.[14]

Whatever Bohr precisely thought about Pauli's remarks on the collapse, he most likely could not have accepted a full quantum description of both object and instrument.[15] For, as already noted, Bohr would later insist on a classical description of

---

[11] I have inserted the word "[But]" in accordance with the original German version of the letter, in which the relevant sentence begins "Um aber" [But to]. The original letter text is in Kalckar (1985, p. 432).

[12] This also means that taking the partial trace cannot constitute a solution to the measurement problem (see next section for Bohr's view on this problem). According to Bacciagaluppi (2008, p. 281) there is, at this point in time, no awareness of a measurement problem in the modern sense of macroscopic superpositions since the Schrödinger equation is not applied to the interaction between objects and instruments. For a discussion of the historical context and meaning of Pauli's claim that no reductions are needed when all instruments are included in the system, see Bacciagaluppi and Valentini, (2009, p. 162 ff).

[13] However, Pauli may have referred instead to the idea that the theoretical description of the closed system should encompass both the (classical) instrument and the (quantum) object, in which case this is not an example of entanglement between two quantum systems (but rather something like the view of Landau and Lifschitz described below).

[14] Bohr's change of mind with regards to not mentioning the collapse in the published Como lecture may of course also have been influenced by discussions with other participants at the Solvay meeting – not least Einstein, Dirac, Born and Heisenberg. Bacciagaluppi (2008) discusses the view of these physicists at the time, and focuses in particular on the possible changes in Born and Heisenberg's view on the collapse (the notion of which is not mentioned in their joint report on quantum mechanics at the meeting).

[15] Closely related to this issue, Bohr would soon (e.g. in 1935) insist that the placing of the split between the (quantum) object and the (classical) instrument is given by the nature of the problem studied and the experimental context (see also section 3). By contrast, Heisenberg (and perhaps Pauli) thought that the placing of the cut can be moved arbitrarily in the direction of the observer; see Schlosshauer and Camilleri (2008) for references and discussion of this disagreement. However, Bohr may not have been settled on this already in 1927. In the beginning of the Como lecture, he seems to be close to Heisenberg's view when he notes that "…the concept of observation is in so far arbitrary as it depends upon which objects are included in the system to be observed", and that "...for every particular case it is a question of

at least part of the measurement apparatus. Nevertheless, he may well have taken from Pauli that the collapse is merely a formal device which is somehow related to the split between the instrument and the object. This would go some way to explain that Bohr never speaks of the collapse in his published writings. Moreover, Bohr might well have worried that even mentioning the collapse could suggest that we are talking about a physical process – a real collapsing wave – which, in his view, we clearly are not.

Although the collapse is not explicitly mentioned, it does seem to be implied in Bohr's 1949 account of his discussions with Einstein at the Solvay meeting. Thus, Bohr comments on Einstein's "…deep concern over the extent to which causal account in space and time was abandoned in quantum mechanics" (Bohr 1949, p. 41), and he goes on to discuss an example, used by Einstein on this occasion, of a particle passing through a slit and subsequently being registered on a photographic plate. Bohr notes (1949, p. 42):

> The apparent difficulty, in this description, which Einstein felt so acutely, is the fact that, if in the experiment the electron is recorded at one point A of the plate, then it is out of the question of ever observing an effect of this electron at another point (B), although the laws of ordinary wave propagation offer no room for a correlation between two such events.

Thus, *apparently*, we have to do with a wave-to-particle transition, or a physical collapse in which no interference between different components of the wave function is possible after the collapse to a single component has taken place. Indeed, according to the notes of the discussions from the Solvay meeting (Kalckar 1985, p. 102), Einstein's point was that "…the interpretation according to which $|\psi|^2$ expresses the probability that *this* particle [pertaining to an individual process] is situated at a certain place presupposes a very particular [or "entirely peculiar"] mechanism of action at a distance [the collapse] which would prevent the wave continuously distributed in space from acting at *two* places of the screen" (Kalckar 1985, p. 102, my inserts, emphases in original).[16] Furthermore, Einstein notes that this action-at-a-distance difficulty follows from describing the process solely in terms of a (real) Schrödinger wave, and he suggests that the problem can only be circumvented by assuming the particle to be localized also during its propagation. By contrast, on Bohr's view, the difficulty is only apparent as we are not dealing with a real wave representing the electron, and as no causal space-time description of the registration process can be given (see also below).

Returning to the Bohr-Pauli correspondence, the collapse is to my knowledge not further mentioned until 1955. In a discussion of the notion of a 'detached observer' and of whether observations somehow create the results, Pauli again refers to the reduction of wave packets (Favrholdt 1999, p. 564):

> In quantum mechanics …an observation hic et nunc [here and now] changes in general the "state" of the observed system in a way not contained in the mathematically formulated *laws,* which only apply to the automatical time dependence of the state of a *closed* system. I think here

---

convenience at what point the concept of observation involving the quantum postulate with its inherent 'irrationality' is brought in" (Bohr 1928, 54).

[16] The inserted expression "entirely peculiar" is from the published version of the Solvey proceedings, see Bacciagaluppi and Valentini (2009, p. 487) and also (2009, section 7.1) for more discussion of Einstein's remarks. As explained by these authors (2009, p. 476, note a), the Bohr Archive's notes in Kalckar (1985) differ somewhat from what was finally published in the Solvay proceedings.

on the passage to a new phenomenon by observation which is technically taken into account by the socalled "reduction of the wave packets." [My insert, emphasis in original]

(Pauli to Bohr, 15 February, 1955)

Pauli is here referring to the well-known distinction between the deterministic evolution in Schrödinger's equation (the law) and the collapse which he suggests is brought about by the observer. In his response, Bohr writes on this point (Favrholdt 1999, p. 568):

I take it for granted that, as regards the fundamental physical problems which fall within the scope of the present quantum mechanical formalism, we have the same view, but I am afraid that we sometimes use a different terminology.[17] Thus, when speaking of the physical interpretation of the formalism, I consider such details of procedure like "reduction of the wave packets" as integral parts of a consistent scheme conforming with the indivisibility of the phenomena and the essential irreversibility involved in the very concept of observation [or recording].[18] As stressed in the article, it is also in my view very essential that the formalism allows of well defined applications only to closed phenomena, and that in particular the statistical description just in this sense appears as a rational generalization of the strictly deterministic description of classical physics.

(Bohr to Pauli, 2 March, 1955)

Bohr here associates three aspects with the collapse: 1) Indivisibility of the phenomena; 2) the essential irreversibility involved in a recording; and 3) that the quantum formalism is only applicable to closed phenomena (= those whose observation results in definite results, see next section) which are associated with a statistical description. As we saw in the discussion of complementarity above, the first point of indivisibility means that the phenomena are not further analyzable, due to the incontrollable interaction (or dynamic inseparability) between quantum system and apparatus. Thus, we cannot give a precise causal space-time account e.g. of the formation of a dot on the photographic plate in a double slit experiment.

The collapse cannot therefore correspond to a physical process amendable to a causal space-time description (in conformity with the view that the wave function does not represent a physical wave). Calling the collapse a "detail of procedure" again suggests that Bohr took it to be a formal (as opposed to physical) notion in which a superposition is replaced by one of its components. As noted in the above subsection, this replacement is due to the change of context associated with an irreversible recording and not because an observer looks at the system. We can thus say that, for Bohr, the collapse is not physical in the sense of a physical wave (or something else) collapsing at a point. But it is a description – in fact the best, or most complete, description – of something happening, namely the formation of a measurement record (e.g. a dot on a photographic plate).

---

[17] I thank Finn Aaserud for the remark (in private communication) that this is the strongest possible form in which Bohr can say that he deeply disagrees with Pauli on this issue!

[18] In Bohr's next letter to Pauli (from March 25, 1955), he explains that by 'observation' he is not referring to any active role of the observer in bringing about the measurement outcome: "Indeed, I really think that our divergency is more related to the use of the word observation itself, with which I simply understand a recording which is unambiguously communicable in common language without requiring any, further creative treatment" (Favrholdt 1999, p. 575).

A mathematical model of the collapse, which may roughly capture the spirit of Bohr's idea in formal terms, can be found in Landau and Lifshitz (1981, pp. 21-24). Landau and Lifshitz (p. 22) speak about a change or collapse of the total system (object × apparatus) upon a measurement. As they point out, this total system is *not* taken to be a quantum system, but rather a combination of a quantum and a classical system. Formally, Landau and Lifshitz assign a 'quasi-classical' wave function to the apparatus, which allows for the fact that the apparatus is always – before, during and after measurement – in a definite state. Nevertheless, as already mentioned (see also below), Bohr argued that a part of the measurement apparatus may be included in the quantum system, in which case a measurement on this part corresponds to a collapse of an entangled state of two quantum systems.[19]

## 3. The necessity of the classical

I now return to Schrödinger's question noted in the introduction. Why did Bohr insist on a classical description of the experimental arrangement and the results of observations? In order to answer this question, one should first be clear about what precisely Bohr understood by classical concepts and descriptions. Although there has been some debate on this issue, commentators usually take Bohr to refer straightforwardly to the concepts and descriptions used in classical physics (see e.g. Bokulich and Bokulich 2005). This standard idea is supported already in the Como lecture where Bohr notes that there is "a fundamental contrast between the quantum of action and the classical concepts" (1928, p. 57), and that "the limitation in the classical concepts [are] expressed through [the uncertainty] relation" (1928, p. 60). Classical concepts are therefore those in which Planck's constant do not enter, e.g. the ordinary, commuting, position and momentum variables from classical physics. Correspondingly, a classical description is one in terms of classical concepts, and one in which there is no restriction to the simultaneous use of pairs of such concepts (as opposed to complementarity).[20]

As illustrated by the first quote in the introduction, it is clear that Bohr took the use of classical concepts and descriptions to be a necessary condition for unambiguous communication. While many authors have focused on Bohr's emphasis on the role of language, Howard (1994, p. 208 ff.) points out that Bohr's precise argument regarding this role is more subtle than usually assumed. According to Howard, the best interpretation of Bohr on this point is that classical concepts must be used "…because they embody the instrument-object separability assumption" (1994, p. 209). This separability is necessary for the very idea of a measurement since it is what "…enable us to say that *this definite object* possesses *this definite property*" (1994, p. 209, emphasis in original). I think Howard is right concerning the close relation for Bohr between an objective (and unambiguous) description and the separability or distinction between object and instrument. But, as already noted, I do not agree with Howard that Bohr took this separability assumption to be based on a factual entanglement between two quantum systems (object and instrument). This is related to two additional but often overlooked arguments given by Bohr for the necessity of classical descriptions. These could be called, respectively, *closedness* and *reference system*. The first argument also contains what I think must be Bohr's answer to the standard measurement problem in quantum mechanics.

---

[19] In general, a measurement for Bohr can refer either to a fixation of the external (experimental) conditions – which implies a fixation of the initial state of the atomic system – or to the obtainment of a recording. Bohr held that we have a well-defined quantum phenomenon only when both kinds of measurements have been made (Bohr 1938, p. 101). The need for such a combination of measurements is strikingly illustrated in modern 'which-path' and 'quantum eraser' experiments. For instance, one can in a double-slit like experiment postpone the decision between getting interference effects or path information (by fixing the external conditions) until *after* the recording of the particle; see e.g. Herzog et al. (1995).

[20] Against this standard account, Howard's (1994, 2004) reconstruction of Bohr takes classical descriptions to be descriptions in which one (in a given experimental context) denies the entanglement between, and so assumes the separability of, the object studied and the measurement apparatus. Some of my reasons for disagreeing with Howard's reconstruction have been given above, and others will become apparent below.

Let me start with the closedness argument and the measurement problem. In his published writings Bohr never discusses this problem, originally addressed by von Neumann in 1932, in the usual terms of expected superpositions of the states of the measurement apparatus.[21] Nevertheless, Bohr was familiar with von Neumann's account (on which he briefly comments in Bohr 1938, p. 120), and I think it is most likely that Bohr was also aware that his own view constituted a solution, or rather dissolution, of the measurement problem. For example, Bohr (1954, p. 73) says:

> …every atomic phenomenon is closed in the sense that its observation is based on registrations obtained by means of suitable amplifications devices with irreversible functioning such as, for example, permanent marks on a photographic plate … [T]he quantum-mechanical formalism permits well-defined applications referring only to such closed phenomena.

Thus, the quantum mechanical formalism can only be applied in a well-defined manner to phenomena whose observation results in definite and irreversible outcomes (call this closedness).[22] But the point of the measurement problem is precisely that a pure quantum mechanical treatment of both the apparatus and the studied object cannot give rise to such definite outcomes. So Bohr's point can be seen as a solution to (or dissolution of) the measurement problem: If the apparatus registering the outcome is classical, so that it has a well-defined state at any time, then there is no problem with macroscopic superpositions at the end of measurements.

A more precise version of this argument of closedness (involving the measurement problem) seems to be hinted at in an article from 1958. Bohr is first, once again, pointing out that (1958, p.3):

> …the description of the experimental arrangement and the recording of observations must be given in plain language, suitably refined by the usual [classical] physical terminology. This is a simple logical demand, since by the word "experiment" we can only mean a procedure regarding which we are able to communicate to others what we have done and what we have learnt. [My insert]

Here and elsewhere Bohr seems to presuppose that experiments have unambiguous (definite) and communicable results and, indeed, this must be so for the experimental method to make sense. Add to this the premise that *if* instruments were quantum, then there would not be any definite results. This premise would be familiar to Bohr not only from Von Neumann's work (and his projection or collapse postulate) but also from what

---

[21] For a simple introduction to the measurement problem, see e.g. Zinkernagel (2011, section 2.1).
[22] As we saw in section 2.3, irreversibility and definiteness form part of Bohr's view on the collapse.

Pauli noted in his 1927 letter quoted above; namely that if we do not make an instrument/object split, we cannot "describe the results of observation theoretically at all".[23] With these premises in place, we can understand Bohr's insistence on "…the introduction of a *fundamental distinction between the measuring apparatus and the objects under investigation*" which is "a direct consequence of the necessity of accounting for the functions of the measuring instruments in purely classical terms, excluding in principle any regard to the quantum of action" (Bohr 1958, p. 3, emphasis in original). Contrary to the standard measurement problem (which starts from the theory), Bohr can thus be seen to take the unambiguous and definite results of measurements as the interpretational starting point. Seeing the necessity of the classical in light of this argument of closedness is supported also by the end of Bohr (1958, p. 6), in which he mentions "…the emphasis on permanent recordings under well-defined experimental conditions as the basis for a consistent interpretation of the quantum formalism…".

The argument of closedness and the associated dissolution of the measurement problem still leave open precisely how and where the distinction between the classical and the quantum should be made. To clarify this it is helpful to consider again the double slit experiment. In a section entitled "the observation problem in quantum theory", which includes a discussion of this experiment, Bohr writes (1938, p. 104):

> In the system to which the quantum mechanical formalism is applied, it is of course possible to include any *intermediate auxiliary agency* employed in the measuring process. [My emphasis]

An example of such an intermediate auxiliary agency is the spring-suspended diaphragm with one slit, which may be used for a momentum control (and hence particle path determination), in the double slit experiment (see also Bohr 1949, p. 50). Bohr continues the former quote (1938, p. 104):

> Since, however, all those properties of such agencies which, according to the aim of the measurement, have to be compared with corresponding properties of the object, must be described on classical lines, their quantum mechanical treatment will for this purpose be essentially equivalent with a classical description. The question of eventually including such agencies within the system under investigation is thus purely a matter of practical convenience…

In the double-slit case, what I think Bohr here has in mind is that we can determine the momentum of the suspended diaphragm just as in classical physics; e.g. by employing momentum conservation in a collision process between the diaphragm and some test body (see Bohr 1935, p. 698). Such a determination can be made with any desired precision, although there will be a corresponding uncertainty in the position of the diaphragm and of the particle when the particle passes. Therefore, as regards the momentum determination, there is no difference (but rather an "essential equivalence") between a classical and a quantum description of the diaphragm. Moreover, while the uncertainty in the position of the diaphragm is important for the quantum phenomena (as no interference pattern results when the path is determined), it is of course a small uncertainty in comparison to the dimensions of the diaphragm. This is in accordance with the pragmatic point that quantum effects can usually be ignored for macroscopic bodies. These observations suggest that the distinction between classical and quantum coincides with that between measuring instruments and objects (where the latter may include *parts* of the apparatus, such as the spring-suspended diaphragm).[24]

---

[23] Apart from Pauli's letter, Bohr's knowledge of this premise can also be inferred from Heisenberg's remark from the 1927 Solvay meeting: "In quantum mechanics, as professor Bohr has displayed, observation plays quite a peculiar role. One may treat the whole world as one mechanical system, but then only a mathematical problem remains while the access to observation is closed off. To get an observation, one must therefore cut out a partial system somewhere from the world, and one must make "statements" or "observations" just about this particular system" (Kalckar 1985, p. 141).

Notwithstanding the necessary distinction between the classical and the quantum, Bohr stresses that the classical description of objects is strictly speaking valid only as an approximation in the limit where one can ignore the quantum of action. In some sense then, the description of any object depends on the quantum of action. However, this does not imply that Bohr's requirement of a classical description of the measuring device is merely a pragmatic or epistemological arrangement, and that the device is really a quantum mechanical system.[25] For the application of quantum mechanics to a system depends on disregarding – not just in practice but also in principle – the quantum of action for some *other* system. There are at least two reasons for this, and they have to do, respectively, with the already mentioned measurement problem and with the reference system argument for the necessity of classical concepts and descriptions, to which we now turn.

Bohr insisted that when an object, e.g. a part of a measuring apparatus, is treated quantum mechanically, there must always then be a different system which is treated classically. As we saw in section 2.2, this is formally reflected in the contextuality of the wave function, that is, that such a wave function can be ascribed to a system only in a given, classically described, context. More specifically regarding reference systems, Bohr concludes the former quote on the possibility of including part of the apparatus in the quantum description as follows (1938, p. 104):

> ...in each case some ultimate measuring instruments, like the scales and clocks which determine the frame of space-time coordination – on which, in the last resort, even the definitions of momentum and energy quantities

---

[24] Howard (1994) dismisses this common reading, which he calls the "coincidence interpretation", of Bohr. In its place, Howard (1994, p. 217 ff.) interprets the mentioned "essential equivalence" between classical and quantum descriptions in terms of the distinction between mixed and pure states in quantum mechanics (which Howard takes to correspond to the classical-quantum distinction). However, apart from the dismissal of the coincidence interpretation being in tension e.g. with Bohr's remarks in his response to EPR (see in particular 1935, p. 701), Howard's reconstruction is difficult to square with Bohr's demand that some parts of the measurement context must be described *entirely* on classical lines (note that a mixed quantum state is still a quantum state). Furthermore, as mentioned also by Schlosshauer and Camilleri (2008, p. 19), on Howard's account Bohr ultimately does not solve the measurement problem.

[25] Thus, in my view, Bohr did not subscribe to the idea that a quantum description of the measurement apparatus is avoided merely 'for all practical purposes' due to suppression of quantum effects for macroscopic bodies. Ideas such as this, nowadays pursued within the framework of decoherence, have been attributed to Bohr e.g. by Rosenfeld and Murdoch (see Murdoch 1987, p. 117). Nevertheless, as also pointed out in a very recent paper by Camilleri and Schlosshauer (2015), Bohr often referred to this pragmatic argument about the heaviness of the apparatus suppressing quantum effects. Camilleri and Schlosshauer take this as an indication for the view that Bohr's insistence on a classical description of the apparatus is merely epistemological (related to a "…functional account of the experiment as a means of acquiring empirical knowledge", (2015, p. 79)), while the apparatus ontologically is a quantum system. However, as noted above, this view is incompatible with what I believe was Bohr's response to the measurement problem which, on the assumption that the apparatus is ontologically quantum, remains unresolved (see also Zinkernagel 2015 and discussion below).

rest – must always be described *entirely* on classical lines, and consequently kept outside the system subject to quantum mechanical treatment. [My emphasis]

Bohr is here saying that the definitions of momentum and energy quantities (and, we may assume, spatial and temporal quantities) depend on a classically described reference frame. A way to understand Bohr's point is that we need such a reference frame to make sense of, say, the position or momentum of an electron (in order to establish with respect to *what* the electron has a position or momentum). This reading is supported by Bohr's response to Einstein, Podolsky and Rosen (Bohr 1935, p. 699), in which he says (about measurements on one of two previously interacting particles):

…to measure the position of one of the particles can mean nothing else than to establish a correlation between its behavior and some instrument rigidly fixed to the support which defines the space frame of reference.

As Dickson points out in connection with this quote, Bohr is not here endorsing operationalism (which would imply that position is defined in terms of a procedure for its measurement) but is rather claiming that "...a well-defined frame of reference is crucially a part of the notion of position." (Dickson 2002, p. 269). Given that there is no absolute reference frame, or at least none which is empirically accessible, Bohr's point in the quote can therefore be taken to be simply that any position (or momentum) adscription to an object must be specified with respect to a reference frame.

The reference frame argument for classicality follows when it is noted that, by definition, a reference frame has a well-defined position and state of motion (momentum). Thus the reference frame is not subject to any Heisenberg uncertainty, and it is in this sense (and in this context) entirely classical. This does not exclude that any given reference system could itself be treated quantum mechanically, but we would then need another – classically described – reference system e.g. to ascribe position (or uncertainty in position) to the former.[26]

Both Bohr's dissolution of the measurement problem and the reference frame argument suggest the view that any system may in principle be seen as and treated quantum mechanically, but not all systems can be seen as and treated in this way simultaneously.[27] Bohr rejected in this sense what may be called 'quantum fundamentalism'; the idea that all systems are essentially of a quantum nature and are ultimately describable as such (see also Rugh and Zinkernagel 2005, and Zinkernagel 2015). Instead, Bohr's view seems to have been a form of ontological contextualism: whether an object is quantum or classical depends on the context.

---

[26] See also Dickson (2004) who argues that this requirement of a classical reference frame is crucial in Bohr's reply to Einstein, Podolsky and Rosen regarding the incompleteness of quantum mechanics.

[27] From this follows, for instance, that a detailed investigation could in principle reveal a bit of jiggling of the center of mass of any macroscopic system, in accordance with the uncertainty relation for position and momentum (see also Bohr 1928, p. 67). But it is an open question whether the claim can be sustained for all physical properties, that is, whether a quantum treatment can be given for *all* degrees of freedom of a physical system (see Rugh and Zinkernagel 2016). The application of quantum theory needs clearly specified conditions, including well-defined state vectors, Hilbert spaces, classical contexts, etc. This point in particular questions whether properties such as the vitality of a cat can be given a quantum treatment (that is, whether often used expressions such as "|dead cat>" at all correspond to some pure state in a Hilbert space with well-defined properties), see also Dickson (2007, p. 298). In any case, Bohr seems to exclude the possibility of a quantum treatment of living systems when he comments (1957, p. 100): "...an account, exhaustive in the sense of quantum physics, of all the continually exchanged atoms in the organism not only is infeasible but would obviously require observational conditions incompatible with the display of life".

## 4. Relationship between the classical and the quantum

Quantum mechanics is for Bohr a rational generalization of classical mechanics. This idea is sometimes understood as a consequence of the fact that formal expressions of classical mechanics can be derived from quantum mechanics, in the limit where all actions are large compared to Planck's constant. While this common idea was part of Bohr's thinking, the classical-quantum relation was by no means a way to eliminate (or reduce away) classical mechanics in some limit. Rather, a mutual dependence between the two theories obtains (Bohr 1938, p. 96):

> Indeed, as adequate as the quantum postulates are in the phenomenological description of the atomic reactions, as indispensable are the basic concepts of [classical] mechanics and electrodynamics for the specification of atomic structures and for the definition of fundamental properties of the agencies with which they react. Far from being a temporary compromise in this dilemma, the recourse to essentially statistical considerations is our only conceivable means of arriving at a generalization of the customary way of description sufficiently wide to account for the features of individuality expressed by the quantum postulates and reducing to classical theory in the limiting case where all actions involved in the analysis of the phenomena are large compared with a single quantum. In the search for the formulation of such a generalization, our only guide has just been the so called correspondence argument, which gives expression for the exigency of upholding the use of classical concepts to the largest possible extent compatible with the quantum postulates. [My insert]

Besides the mutual dependence between classical and quantum theory, Bohr here refers to the correspondence argument or principle, and the idea of retaining the classical concepts as far as possible.[28] Given these connection points between the classical and the quantum, Bohr would most likely have agreed with the succinct expression of Landau and Lifshitz (1981, p. 3):

> Quantum mechanics occupies a very unusual place among physical theories: it contains classical mechanics as a limiting case, yet at the same time it requires this limiting case for its own formulation.

Bohr's insistence of the classical description of the apparatus is one way to bring out this requirement. We have already discussed Bohr's view on this issue, which has of course been noted by many authors, e.g. Tanona (2004, p. 690): "[T]he application of the [quantum] theory to any particular experimental arrangement depends on conceptual apparatus external to the theory—i.e., it depends on classical physics."

---

[28] The correspondence argument may be seen as an instance of the correspondence principle, at least in the generalized version also discussed by Heisenberg (see Falkenburg 2007, p. 190). See also Bokulich and Bokulich (2005, p. 349) who in this connection cite Bohr's idea that a rational generalization of classical physics "…would permit the harmonious incorporation of the quantum of action" (Bohr 1958, p.2).

However, quantum theory is surely relevant also for the description of phenomena not directly related to experimental arrangements. To see the generality of Bohr's view of the classical-quantum relationship, it is therefore helpful to consider one of the few examples of a quantum phenomenon Bohr mentioned, in which no direct reference to a measuring apparatus is made. The example in question is that the stability of atoms, and therefore of matter, depends on the quantum of action or the 'quantal laws'. This dependence is mentioned on several occasions in Bohr's writings, e.g.:

> …it is indeed only the existence of the quantum of action which prevents the fusion of the electrons and the nucleus into a neutral massive corpuscle of practically infinitesimal extension. (Bohr 1937, p. 17-18)

> …we meet in experimental evidence concerning atomic particles with regularities of a novel type, incompatible with deterministic analysis. These quantal laws determine the peculiar stability and reactions of atomic systems, and are thus ultimately responsible for the properties of matter on which our means of observation depend. (Bohr 1958, p. 2)

Regarding the stability of the atom, Bohr did not, to my knowledge, discuss precisely how classical considerations enter the picture,[29] and I am also unaware of previous discussions of this issue in Bohr scholarship. But I think the mutual dependence between quantum and classical mechanics in this case may perhaps be understood as follows. Heisenberg's uncertainty relation can, at least heuristically, explain why the electron does not spiral in to, and remain at, the center of the atomic nucleus. If an electron could remain at rest at the center of the atom, then both its momentum and its position would be well determined, in clear violation of Heisenberg's uncertainty relation. But this explanation makes sense only in a context where the nucleus is assumed to be at a specific location (at which the electron then cannot remain at rest). In this argument, the nucleus therefore defines a frame of reference which, by definition, is not subject to uncertainty and can thus be seen as classical.[30]

This does not mean that one *must* treat the nucleus classically in order to explain the stability of e.g. the hydrogen atom. The stability may also be accounted for by treating the electron and nucleus (proton) as a quantum two-body system. One can then solve the Schrödinger equation for the coupled quantum system, and find the mean value of the separation to be larger than zero. However, such an approach will again imply reference to some external classical context in order to specify the meaning of the mean or expectation value in terms of expected (definite) outcomes of position measurements.[31] This is consistent with Bohr's point, cited in section 3, that we always need a (classical described) reference frame which is "…kept outside the system subject

---

[29] Although he did indicate that the role of Planck's constant in the stability and functioning of the measurement apparatus, did not change his general outlook on the quantum mechanical description (Bohr 1949, p. 51).

[30] Though Heisenberg uncertainty is often invoked to explain the stability of the hydrogen atom, it is strictly speaking insufficient. Moreover, for many-electron atoms the Pauli exclusion principle also comes into play. It remains true, however, that a basic feature leading to atomic stability is that a sharp localization of an electron (close to or at the nucleus) will increase the kinetic energy (and momentum) without bound; see Lieb (1976) for a discussion of these issues.

[31] I thank Carlo Rovelli (private communication) for bringing my attention to these points on the two-body approach to atomic stability.

to quantum mechanical treatment".[32]

While a nucleus *may* therefore be treated classically, a nucleus, a complete atom or a molecule may in many other contexts be considered and described quantum mechanically. But it requires in all cases, according to Bohr, that parts of the (experimental) context is considered and described in classical terms. That even objects like nuclei can be considered to be either quantum or classical, depending on the context, illustrates once more that the distinction between the quantum and the classical is flexible.

## 5. Outlook

After almost 90 years of discussion, there is still no consensus regarding which among the various alternatives is the better interpretation of quantum mechanics. Indeed, it seems to be widely recognized that any interpretation of the theory will involve some mystery, weirdness or, at least, problematic aspects.

Bohr's interpretation of quantum mechanics involves the necessity of describing in classical terms *some* parts of any system exhibiting quantum effects, though *which* parts may vary from one context to another. I have argued above that this contextual divide between the classical and the quantum forms an essential part of Bohr's view on the wave function, the collapse, and the measurement problem. While I believe Bohr's position to be both coherent and convincing, it does not mean that there is no mystery left. As we have seen, a central ingredient in Bohr's account of the collapse is the indivisibility or un-analyzability of the measurement process (by which, somehow, the object interacts with the instruments to produce a permanent mark in the latter). This is where mystery enters Bohr's account, although it is a mystery with a name and a physical origin (see below). I think this is a perfectly respectable aspect of Bohr's ideas, not least given that quantum phenomena – such as the double slit interference pattern – indeed have a mysterious air. But even if one disagrees, it should be recalled that other interpretations substitute the mystery for other types of weirdness at this point (e.g. a splitting of worlds for Everettians, or the effect of a non-local quantum potential and/or in-principle unmeasureable definite particle positions for Bohmians).

Bohr was no instrumentalist and no mystic. He emphasized that there is a physical reason for the indivisibility; namely the quantum of action or Planck's constant (the name of the mystery!). During the measurement process, the quantum of action induces an incontrollable interaction between the measurement apparatus and the object. This leads to complementarity which implies that we can *either* use space-time descriptions *or* a causal description but not both. And this again means that we cannot pinpoint a causal mechanism in spacetime which ultimately explains why a particular outcome occurs. Of course, nothing prevents us from keep investigating and reveal new features of the measurement process.[33] But, if Bohr's view on quantum mechanics is

---

[32] A similar point is made by Dickson (2004, p. 203) in a discussion of the use of relational coordinates (using just the distance between the electron and the proton) in the determination of the energy levels of hydrogen. Dickson observes that the use of relational coordinates depends on first separating out and then ignoring the center-of-mass coordinates for the combined system. But such center-of-mass coordinates are strictly speaking only defined in relation to some (classical described) reference frame (see also Rugh and Zinkernagel 2005, p. 59).

[33] A case in point is the important phenomenon of decoherence which helps to explain why no quantum effects is seen on the ordinary macroscopic scale. However, as is well known, decoherence does not by itself constitute an adequate response either to the measurement problem or to the more general question

right, then a full microphysical space-time picture of this process will always remain elusive.

Let me conclude with a few remarks on the continued relevance of Bohr's interpretation, not only as a serious candidate for understanding quantum mechanics, but also with respect to quantum ideas in fundamental physics and cosmology. Regarding the former, some approaches to quantum gravity (e.g. Loop Quantum Gravity), assumes the Bohrian view that not everything can be treated quantum mechanically simultaneously. In particular, it may only be possible to address the quantum structure of spacetime (in a very small region) when a classical structure of spacetime (in a large region containing the small one) is presupposed.[34] Regarding modern cosmology, a key issue is the supposed transition between quantum fluctuations and classical density perturbations during an assumed inflationary phase in the very early universe. This transition is believed to lie at the origin of structure formation in the universe. However, the idea is faced with what one may call the 'cosmic measurement problem'; the problem of how to get classical structures from quantum constituents in a cosmological context. Also in this case, it may be argued – in consistency with Bohr's point of view – that any account of the developments in the very early universe will depend on assuming part of the total system to be classical. In particular, such classicality may be needed in order to secure a physically meaningful concept of time, which is a precondition for addressing issues about the 'early' universe. Thus, as argued in Rugh and Zinkernagel (2011, 2013), a well-defined cosmic time concept is closely related to a classically described motion of matter in the universe.

Bohr's interpretation of quantum mechanics has sometimes been accused of putting an end to explanation, or to operate with an *a priori* classical domain which makes it unsuitable for cosmology and theories at the frontiers of fundamental physics. I hope the brief remarks above are sufficient to indicate that such assessments are mistaken.

**Acknowledgements**
I would like to thank Albert Solé for inviting me to a stimulating workshop in Barcelona in 2013 on "Space-time and the wave function". I am grateful to the participants at the workshop, not least Jeff Barrett and Carlo Rovelli, for comments and discussion on some of the ideas presented here. I also thank the anonymous referees for helpful comments, as well as Svend E. Rugh for comments and many discussions on Bohr and quantum theory over the years. Finally, I thank the Spanish Ministry of Science and Innovation (Project FFI2011-29834-C03-02) for financial support.

---

of the relationship between the classical and the quantum (see Schlosshauer 2007 and also e.g. Zinkernagel 2011 and Tanona 2013 for further discussion and references).

[34] See also Zinkernagel (2006). Carlo Rovelli (private communication) agrees with this view and explains that in Loop Quantum Gravity, the quantum behavior of a small region of spacetime is studied "…in terms of how it affects ('is measured') by a surrounding large spacetime region, which is treated classical."